\documentstyle[12pt,epsf]{article}
\newcommand{\bra}[1]{\left\langle #1 \right|}         
\newcommand{\ket}[1]{\left| #1 \right\rangle}         
 
\newcommand{\de}[2]{{d #1 \over d #2}}                
\newcommand{\pa}[2]{{\partial #1 \over \partial #2}}  
\newcommand{\papa}[3]{{\partial^2 #1 \over \partial #2 \partial #3}}

\newcommand{\ska}[1]{\left( #1 \right) }            
          
\newcommand{\bka}[1]{\left[ #1 \right] }            

\newcommand{\afootnote}{\renewcommand{\thefootnote}{\alph{footnote}}}
\newcommand{\Afootnote}{\afootnote\footnote}

\begin{document}
\begin{titlepage}
\begin{flushright}
KANAZAWA/98-22
\end{flushright}
\quad\\
\vspace{1.8cm}
\begin{center}
{\Large \bf Non-Perturbative Renormalization Group\\
\vspace{4mm}
and Quantum Tunnelling
\Afootnote{talk given by A.Horikoshi at the Workshop on the Exact
  Renormalization Group (Faro, Portugal, September 1998)}}
\par
\vspace{1.8cm}
{\large 
Ken-Ichi A{\sc oki}
\Afootnote{e-mail : aoki@hep.s.kanazawa-u.ac.jp},
\ Atsushi H{\sc orikoshi}
\Afootnote{e-mail : horikosi@hep.s.kanazawa-u.ac.jp},
\ Masaki T{\sc aniguchi}
\Afootnote{e-mail : taniguti@snc.sony.co.jp} \\
\vspace{2mm}
and
\   Haruhiko T{\sc erao}
\Afootnote{e-mail : terao@hep.s.kanazawa-u.ac.jp}}

\vskip15mm
Institute\ for\ Theoretical\ Physics, Kanazawa\ University,\\
Kakuma-machi\ Kanazawa\ 920-1192,\ Japan
\vskip10mm
%\maketitle
\end{center}

\begin{abstract}
The non-perturbative renormalization group (NPRG) is applied to
analysis of tunnelling in quantum mechanics. The vacuum
energy and the energy gap of anharmonic oscillators are evaluated by
solving the local potential approximated Wegner-Houghton equation (LPA
W-H eqn.). We find that our results are very good in a strong coupling
region, but not in a very weak coupling region, where the dilute gas
instanton calculation works very well. So it seems that in analysis of 
quantum tunnelling, the dilute gas instanton and LPA W-H eqn. play
complementary roles to each other. We also analyze the supersymmetric quantum
mechanics and see if the dynamical supersymmetry (SUSY) breaking is
described by NPRG method. 
\end{abstract} 
\end{titlepage}

%%%%%%%%%%%%%
\section{Introduction}
The theoretical basis of NPRG was formulated by K.G.Wilson in
1970's.\cite{wk} After that, several types of `exact' renormalization group
equations were derived and have been applied to various quantum
systems. Although those equations are exact, we can not
solve them without any approximation in practice. Therefore it is not trivial
that such NPRG analysis can take account of the effects caused by
non-perturbative dynamics even qualitatively. \par
Generally, there are two types of non-perturbative quantities. One
corresponds to summation of all orders of the perturbative series,
which might be related to the Borel resummation.\cite{gz} The other is the 
essential singularity with respect to coupling constant $\lambda$, which has a structure
like $e^{-\frac{1}{\lambda}}$.\cite{co} We are not able to expand this singular 
contribution around the origin of $\lambda$. This singularity is
essential in case of quantum tunnelling. For example, in the
symmetric double well system, there are degenerated two energy levels
at each minima, which are mixed through tunnelling to generate an energy gap $\Delta E \sim e^{-\frac{1}{\lambda}}$. The
exponential factor comes from the free energy of topological
configurations, the instantons. Can NPRG evaluate these non-perturbative
effects in a good manner? The main purpose of our work
is to check
this not only qualitatively but also quantitatively. We carry it out
in quantum mechanical systems by comparing our results with the exact
values given by numerical analysis of
the Schr\"odinger equation, with the perturbative series, and with the
instanton method. The instanton method is a unique analysis of quantum 
tunnelling leading to the exact essential singularity, which is,
however, valid only in a very
weak coupling region. It will turn out
that the instanton method and LPA W-H eqn. are somehow
complementary to each other. \par
Though quantum tunnelling is one of the most striking consequences
of quantum theories, there has been no general-purpose tool to analyze 
it. So if NPRG treats non-perturbative dynamics well, it can be a
powerful new tool for analysis of tunnelling and this work will be the first
touch to attack more complex systems with quantum tunnelling by NPRG.  
\section{The NPRG study of quantum mechanics}
As a primary study, we would like to restrict ourselves to treat
the effective potential. We start with the LPA W-H eqn. for scalar
theories, where we ignore the corrections to the derivative interactions,\cite{wh,ap}
 \begin{eqnarray}
 \pa{\hat{V}_{\rm
 eff}}{\tau}=\left[D-d_{{\varphi}}\hat{\varphi}\pa{}{\hat{\varphi}}\right]\hat{V}_{\rm eff}+{A_D \over 2}
  \log\ska{1+\papa{\hat{V}_{\rm eff}}{\hat{\varphi}}{\hat{\varphi}}} . 
 \end{eqnarray}
Each hatted($~\hat{}~$) variable represents a dimensionless quantity
   with a unit defined by the momentum cut off $\Lambda(\tau)=\rm e^{-\tau}\Lambda_{\rm 0}$, $D$ is the space-time 
   dimension, 
   $d_{\varphi}=\frac{D-2}{2}$ is the canonical dimension of scalar
   field $\varphi$, and $A_D=(2\pi)^{-D}\int d\Omega _{D}$. This is a partial differential equation of 
   the dimensionless effective potential $\hat{V}_{\rm eff}$ with respect
   to $\hat{\varphi}$ and scale parameter $\tau$. \par
Furthermore, we expand $V_{\rm eff}$ as power series of $\varphi$, 
\begin{eqnarray}
 V_{\rm eff} \left( \varphi ;\tau \right) &=&
 \sum_{n=0}^N\frac{a_n(\tau)}{n!}\varphi ^n ,
\end{eqnarray}
which is called the operator expansion. If the results converge as the 
order of truncation $N$
becomes large, we regard them as the solutions of LPA W-H eqn. The partial differential equation is reduced to a set of ordinary differential
equations for dimensionless couplings $\{\hat{a}_n\}$, 
 \begin{eqnarray}
  \de{\hat{a}_0}{\tau}&=&~~~~~~~~D\hat{a}_0+\frac{A_D}{2}\log(1+\hat{a}_2) ,\nonumber \\
 \de{\hat{a}_1}{\tau}&=&~~\frac{D+2}{2}\hat{a}_1+\frac{A_D}{2}\left[\hat{a}_3 \over
 1+\hat{a}_2\right] ,  \nonumber \\
 \de{\hat{a}_2}{\tau}&=&~~~~~~~~~2\hat{a}_2+\frac{A_D}{2}\bka{{\hat{a}_4 \over 1+\hat{a}_2}-
    {\hat{a}_3^2 \over \ska{1+\hat{a}_2}^2}} , \nonumber \\
 \de{\hat{a}_3}{\tau}&=&~~\frac{6-D}{2}\hat{a}_3+\frac{A_D}{2}\bka{{\hat{a}_5 \over 1+\hat{a}_2}-{3\hat{a}_4\hat{a}_3
    \over \ska{1+\hat{a}_2}^2}+{2\hat{a}_3^3 \over
    \ska{1+\hat{a}_2}^3}} , \nonumber \\
 \de{\hat{a}_4}{\tau}&=&(4-D)\hat{a}_4+\frac{A_D}{2}\bka{{\hat{a}_6 \over 1+\hat{a}_2}-{4\hat{a}_5\hat{a}_3
    \over \ska{1+\hat{a}_2}^2}+ {12\hat{a}_4\hat{a}_3^2 \over \ska{1+\hat{a}_2}^3}-{3\hat{a}_4^2
    \over \ska{1+\hat{a}_2}^2}- {6\hat{a}_3^4 \over
    \ska{1+\hat{a}_2}^4}} . \nonumber \\
 && \vspace{3cm} \vdots  
 \end{eqnarray}
In each $\beta$-function (the right-handed side of each equation), the first term represents the canonical scaling and the second term
represents one-loop quantum corrections, respectively. The common
denominator $\frac{1}{1+\hat{a}_2}$ corresponds to the
`propagator'. The constant part of $V_{\rm eff}$, $a_0$, is given by the vacuum
bubble diagrams and is usually ignored. However we will keep it here,
since it plays a crucial role in supersymmetric theories.  \par
Making use of these equations, we can analyze quantum mechanics,
which is $D$=1 real scalar theory with a dynamical variable $x(t)$. We
now evaluate two physical quantities, the vacuum energy $E_0$ and the energy
gap $\Delta E=E_1 -E_0$. The vacuum energy is given by,
 \begin{eqnarray}
   E_0=\bra{\Omega}\hat{H}\ket{\Omega} = V_{\rm
eff} \left. \right|_{x =<x>} .
 \end{eqnarray}
Namely, the minimum value of
$V_{\rm eff}$ gives us the ground energy of the system. The energy gap 
is obtained through the two point correlation
function,
 \begin{eqnarray}   
\mathop{\rm lim}_{t\to\infty}\bra{\Omega}{\rm T}\hat{x}(t)\hat{x}(0)\ket{\Omega}\propto e^{-(E_1-E_0)t} ,
 \end{eqnarray}
while it is evaluated in the LPA as follows,
 \begin{eqnarray}   
\bra{\Omega}{\rm T}\hat{x}(t)\hat{x}(0)\ket{\Omega}
&\stackrel{\rm LPA}{=}&
\int
\frac{dE}{2\pi}e^{iEt}\frac{1}{E^2+m^2_{\rm eff}}\propto
e^{-m_{\rm eff}t} ,
 \end{eqnarray}
where the effective
mass $m_{\rm eff}$ is the curvature at the potential minimum. Comparing the damping factor as $t$ goes to
infinity, the relation $E_1-E_0=m_{\rm eff}$ follows and we use,
 \begin{eqnarray}   
E_{1}&=&\left.V_{\rm eff}\right|_{x=<x>}+\sqrt{\left.\frac{\partial^2V_{\rm eff}}{\partial
x^2}\right|}_{x=<x>}.
 \end{eqnarray} 
Thus we know the energy
spectrum from the information of effective potential.\par
To show the fundamental procedure of
analysis, we consider the case of
the harmonic oscillator. We evaluate the effective potential $V_{\rm eff}$ by solving the differential
equations for dimensionless couplings as follows,   
\newpage
\vspace{5mm}
\fbox{\parbox{4cm}{initial potential \\ $V_0(x)=a_0+\frac{1}{2}a_2 x^2$}}   
\hspace{1mm}$\Lambda(0)=\Lambda_{\rm 0}$
\vspace{-15mm}
%\hspace{2cm}
\begin{flushright}
$\left( a_0~ ,~a_2\right)\longrightarrow \left( \hat{a}_0=\frac{a_0}{\Lambda_0}~ ,~\hat{a}_2=\frac{a_2}{\Lambda_0^2}\right)$

$\Downarrow \tau =0$\\
{\begin{tabular}{|l|} \hline
LPA W-H eqn.\\

$ \de{\hat{a}_0}{\tau}=\hat{a}_0+\frac{1}{2\pi}\log(1+\hat{a}_2)$ \\ 
$ \de{\hat{a}_2}{\tau}=2\hat{a}_2  $ \\
%\mbox{} \\
%\mbox{} \\
%\mbox{} \\
%\mbox{} \\ 
\hline
\end{tabular}}

$\Downarrow \tau =\tau _{\rm f}$\\

$\left( a_{0\rm f}=\Lambda_{\rm f} \hat{a}_{0f}~,~a_{2\rm f}=\Lambda^2
_{\rm f} \hat{a}_{2\rm f}\right)\longleftarrow \left
( \hat{a}_{0\rm f}~,~\hat{a}_{2\rm f}\right)$
\end{flushright}
\vspace{-15mm}
\hspace{5mm}
\fbox{\parbox{4cm}{final potential \\ $V_{\rm eff}(x)=a_{0\rm
f}+\frac{1}{2}a_{2\rm f} x^2$}}   
\hspace{1mm}$\Lambda_{\rm f}=\rm e^{-\tau_{\rm f}}\Lambda_{\rm 0}$\\
\par
\vspace{5mm}
\hspace{-7mm}
In this case, we can carry out the above procedure analytically,
\begin{eqnarray}
a_0 (\Lambda _{\rm f})\!\!&=&\!\!a_0(\Lambda_0)+
 \frac{\sqrt{a_2(\Lambda_0)}}{2\pi}\left[\hat{p}\log \frac{1+\hat{p}^2}{\hat{p}^2}+2\tan^{-1}\hat{p}
 \right]^{\hat{p}=\frac{\Lambda_0}{\sqrt{a_2(\Lambda_0)}}}_{\hat{p}=\frac{\Lambda_{\rm f}}{\sqrt{a_2(\Lambda_0)}}} , \\
a_2(\Lambda _{\rm f})\!\!&=&\!\!a_2(\Lambda_0) .
\end{eqnarray}
If we take initial conditions 
$(a_0(\Lambda_0),a_2(\Lambda_0))=(0,m^2)$, then we get 
$(a_0(\Lambda _{\rm f}),a_2(\Lambda _{\rm f}))=(\frac{m}{2},m^2)$
in the limit $\Lambda_0 \to \infty $ , $\Lambda_{\rm f} \to 0$. 
That is, we can evaluate the zero-point energy $\frac{m}{2}$ as a result
of running of $a_0$. The evolution of $a_0$ freezes where the cut off scale becomes less than
the mass scale. 
\par
\begin{figure}[htb]  
\hspace{8mm}
 \parbox{70mm}{
 \epsfxsize=70mm     
 \epsfysize=70mm
  \leavevmode
\epsfbox{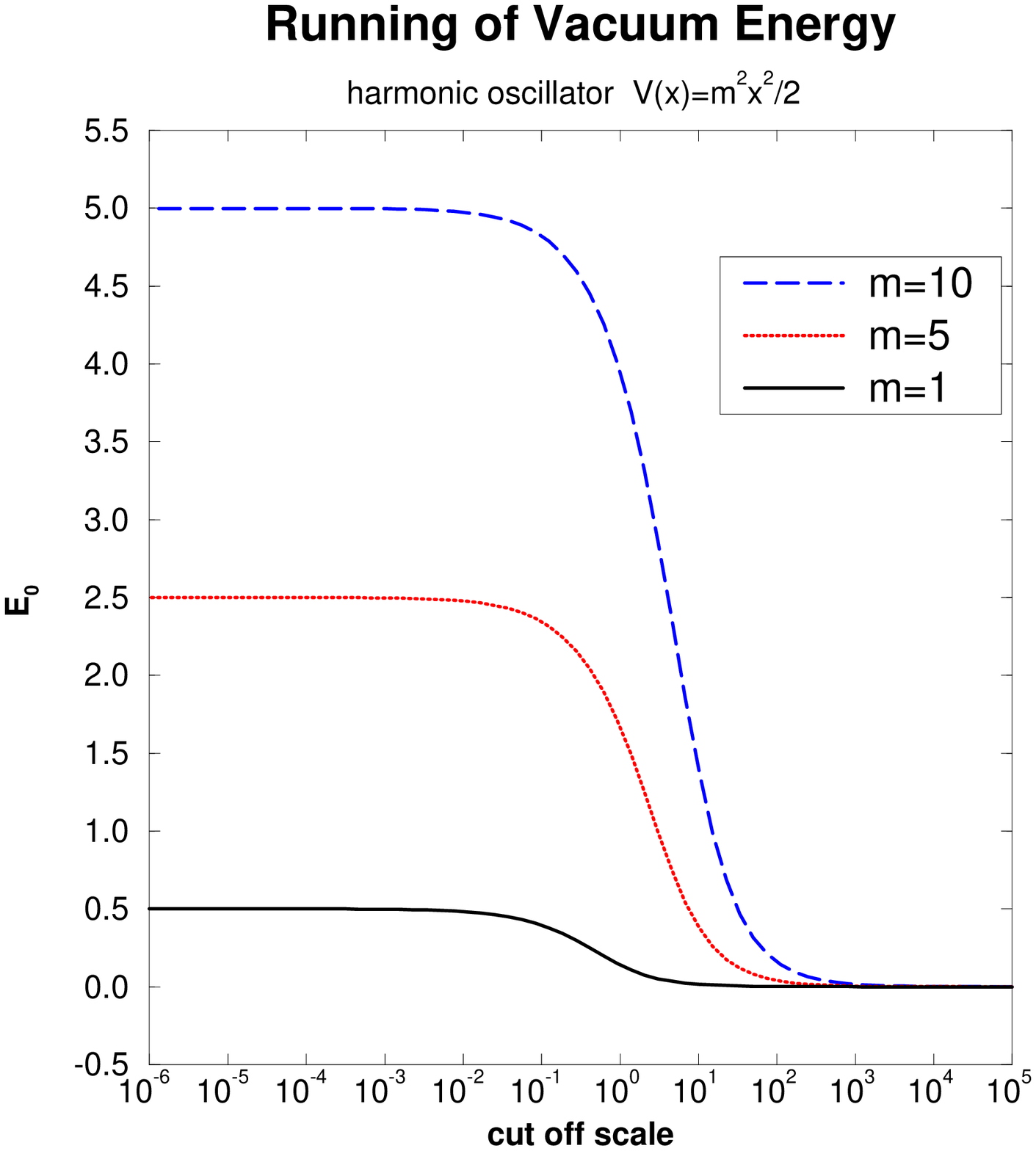}
\vspace{-10mm}
 \caption{Running of vacuum energy}
 \label{fig:hrun}     
 }
\hspace{8mm} 
\parbox{70mm}{
 \epsfxsize=70mm      
 \epsfysize=70mm
 \leavevmode
\epsfbox{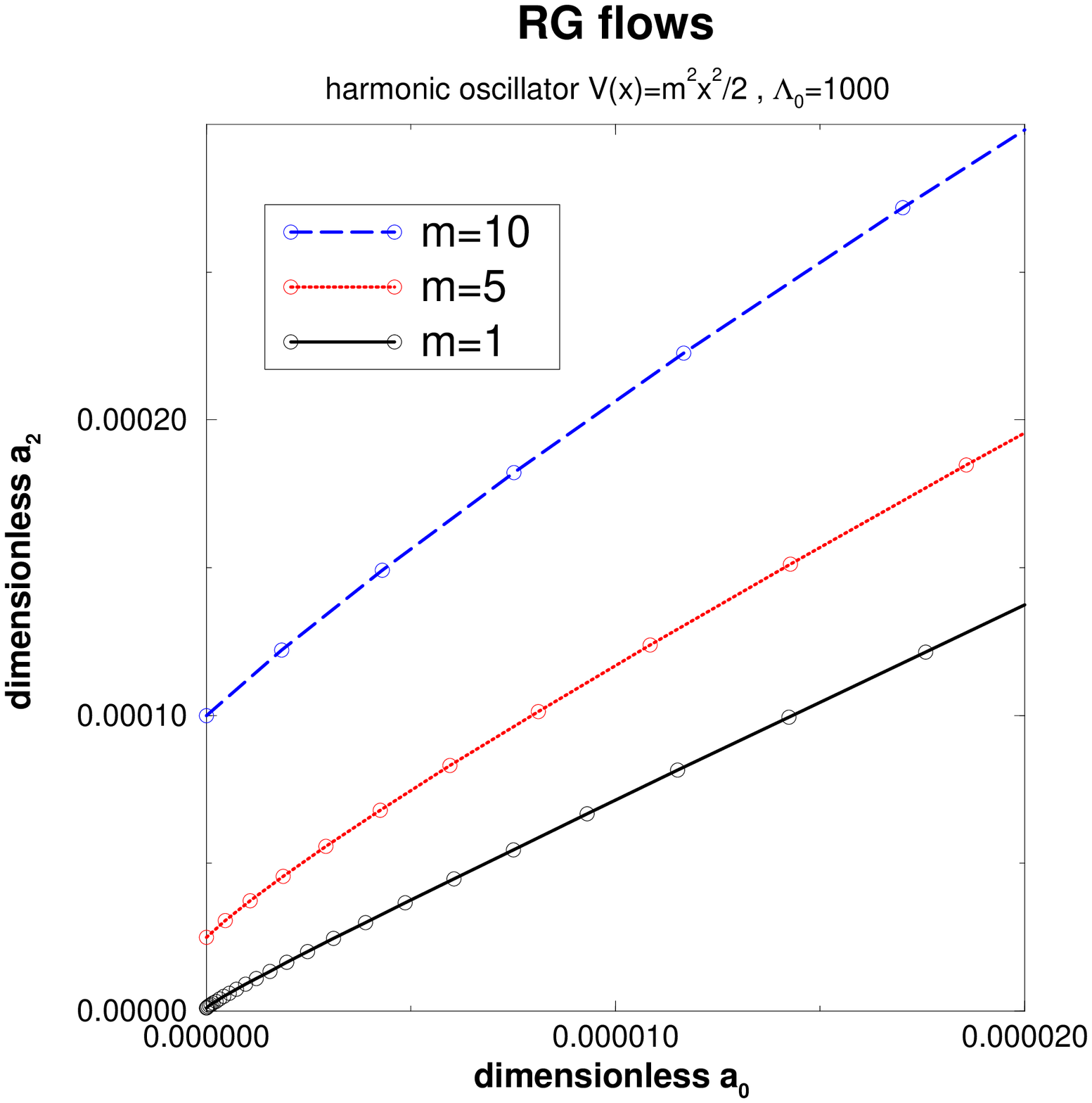}
\vspace{-10mm}
 \caption{Flow diagram}
 \label{fig:hflow} 
 }
\end{figure}
The renormalization group flows are plotted in
Fig.\ref{fig:hrun} and Fig.\ref{fig:hflow}. We see that the momentum region where the quantum corrections are
effective is finite and depends on the mass. It is the decoupling
property, which enables us to get effective couplings as physical
quantities even by numerical calculation within a finite momentum
region.   
\section{Analysis of anharmonic oscillators}
\begin{figure}[htb]
\hspace{8mm}
 \parbox{70mm}{
 \epsfxsize=70mm 
 \epsfysize=70mm
  \leavevmode
\epsfbox{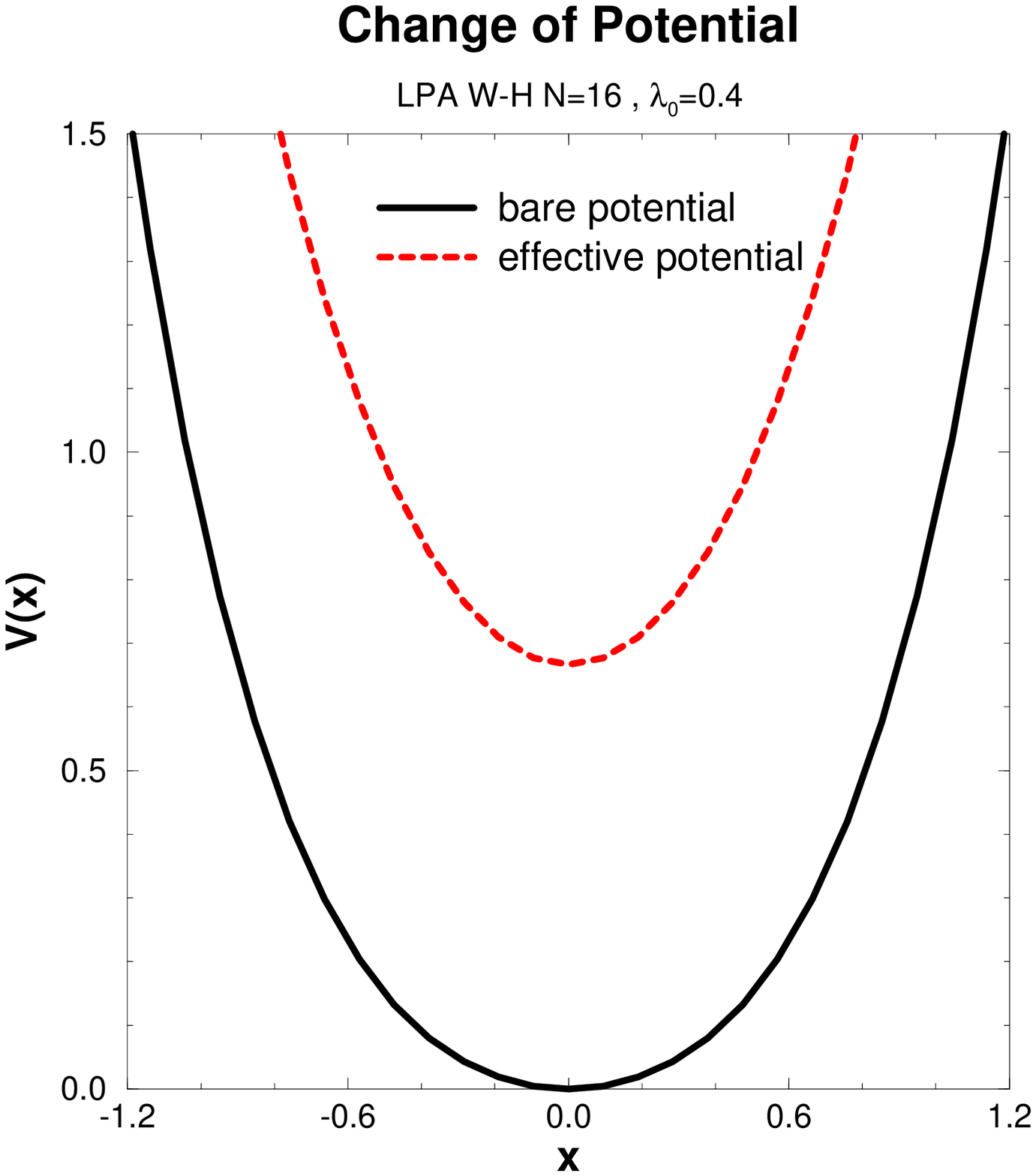}
\vspace{-10mm}
 \caption{Change of potential}
 \label{fig:spote}
 }
\hspace{8mm} 
\parbox{70mm}{
 \epsfxsize=70mm 
 \epsfysize=70mm
 \leavevmode
\epsfbox{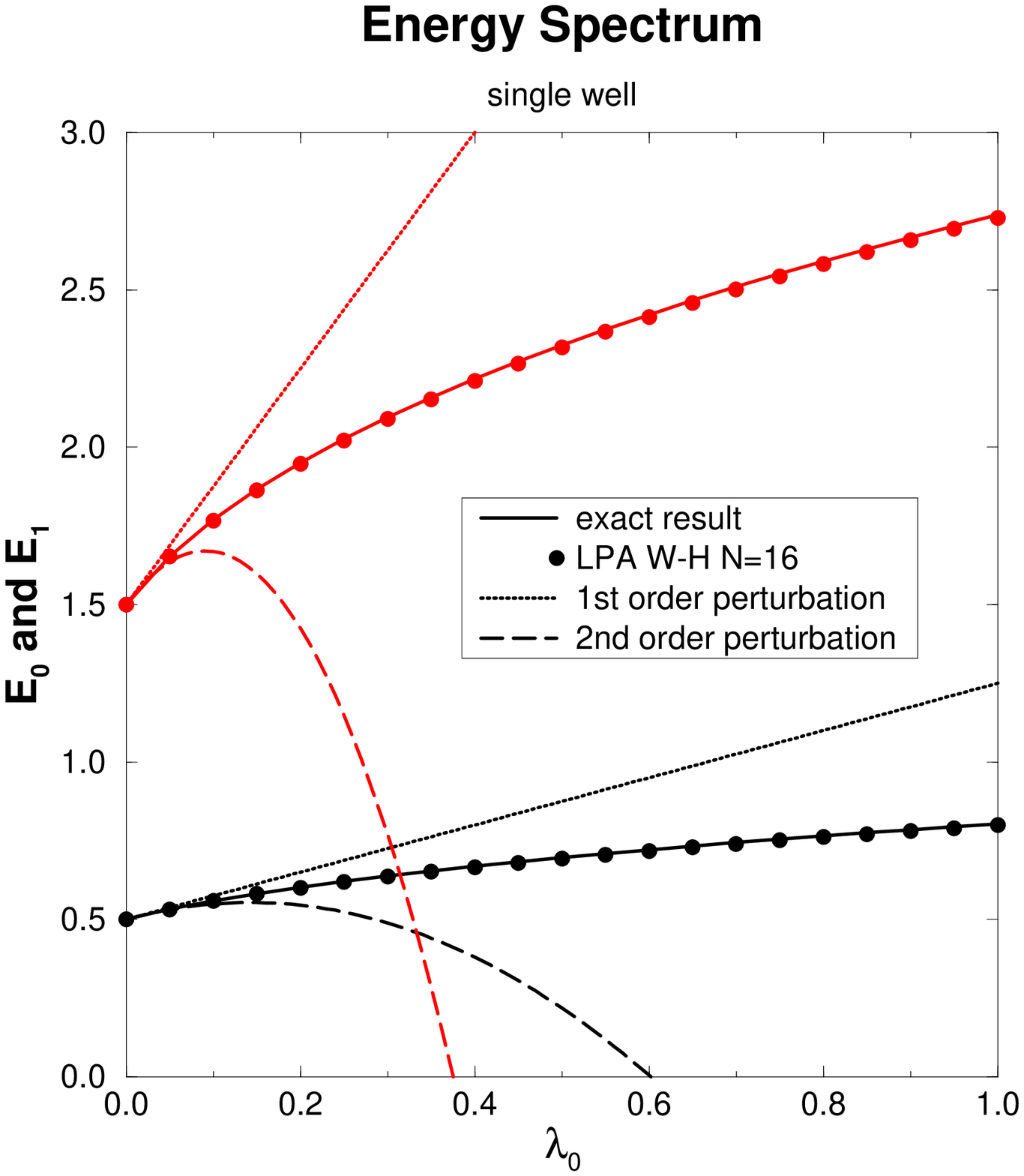}
\vspace{-10mm}
 \caption{Energy spectrum}
 \label{fig:sspe}    
 }
\end{figure}

Now we proceed to analyze quantum mechanics of anharmonic
oscillators. At first, we consider a symmetric single-well potential, 
\begin{eqnarray}
V_0(x) =~~\lambda_0 x^4+\frac{1}{2} x^2 .
\end{eqnarray}
Of course, there is no tunnelling, so our interest is to compare our
NPRG results with the perturbative
series. The corrected $V_{\rm eff}$ is shown in Fig.\ref{fig:spote} and we obtain
the energy spectrum as in Fig.\ref{fig:sspe}.
\par
\par
\hspace{-7mm}
The perturbative series of $E_n$ are the asymptotic series, 
\begin{eqnarray}
E_n\!=\![n+\frac{1}{2}]+\frac{3}{4}\lambda_0 [2n^2+2n+1]
-\frac{1}{8}\lambda_0 ^2 [34n^3+51n^2+59n+21]+\cdots , 
\end{eqnarray}
and shows diverging nature even in the weak coupling region.
Note that the Borel resummation of the perturbative series works fine in this case and gives quantitatively good
values. On the other
hand, even in the lowest order approximation, W-H equation can
evaluate the energy spectrum almost perfectly. Therefore, NPRG could treat all orders of the perturbative
series in a correct manner. \par
\begin{figure}[htb]  
\hspace{8mm}
 \parbox{70mm}{
 \epsfxsize=70mm     
 \epsfysize=70mm
  \leavevmode
\epsfbox{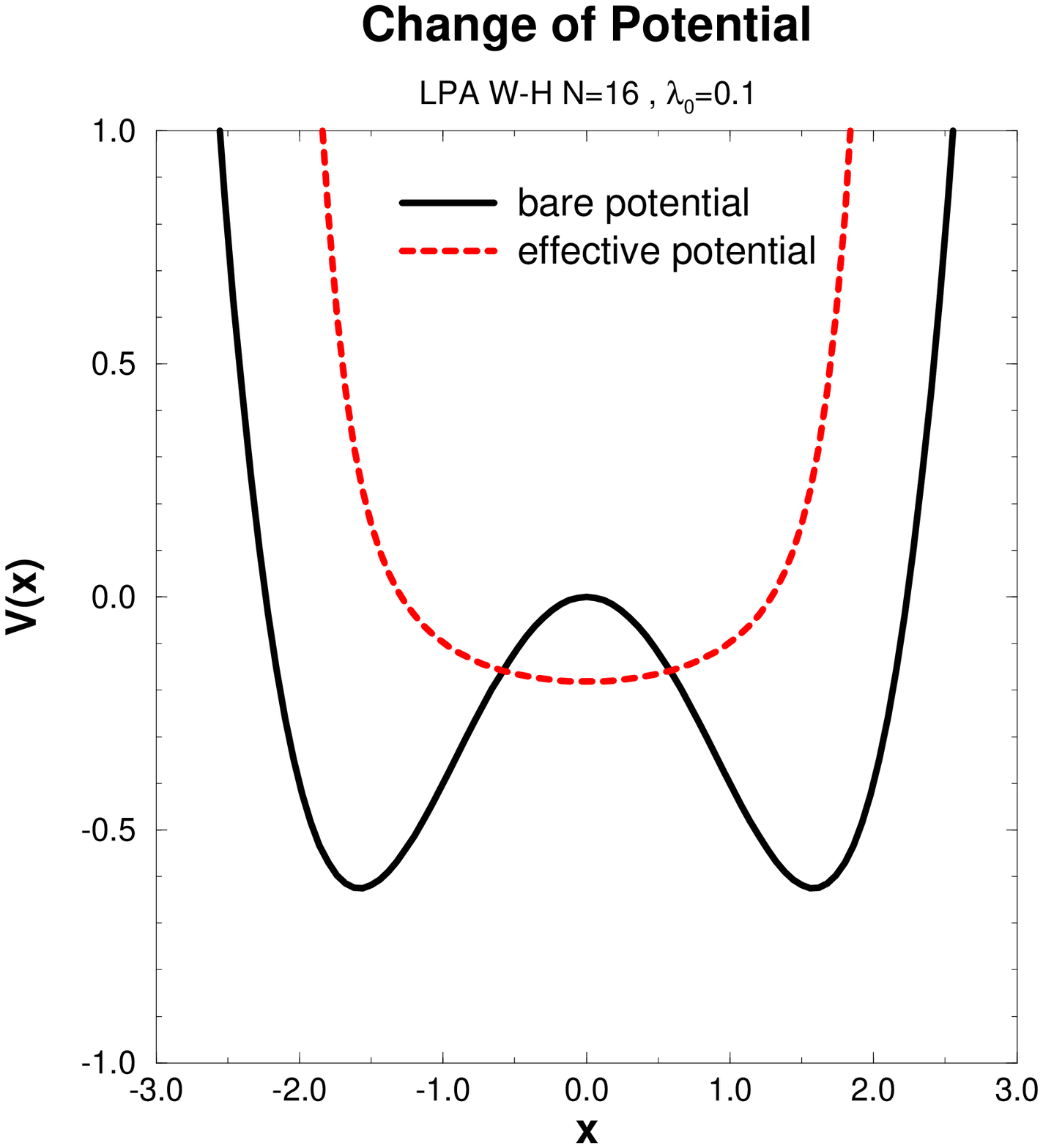}
\vspace{-10mm}
 \caption{Change of potential}
 \label{fig:dpote}   
 }
\hspace{8mm} 
\parbox{70mm}{
 \epsfxsize=70mm     
 \epsfysize=70mm
 \leavevmode
\epsfbox{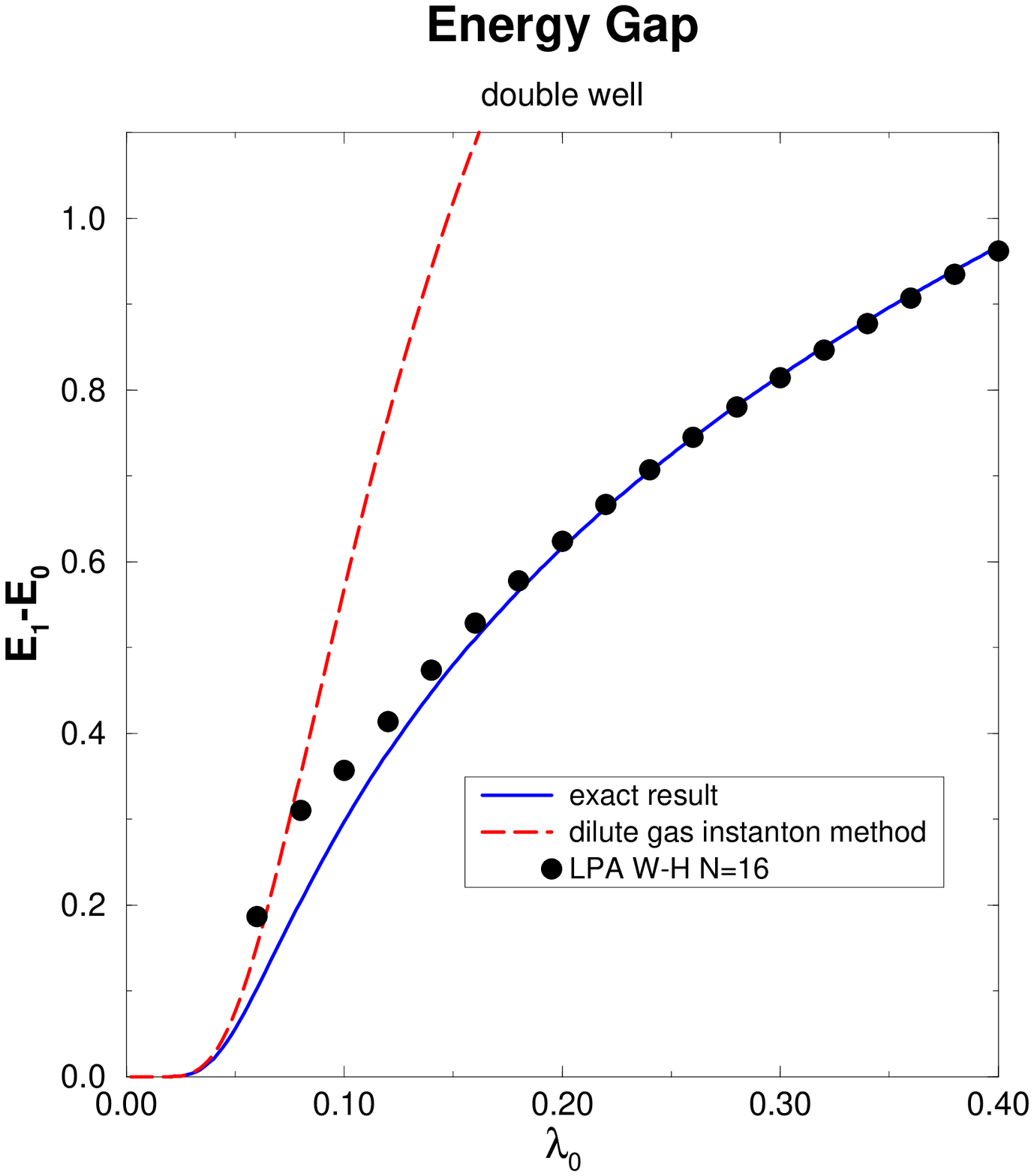}
\vspace{-10mm}
 \caption{Energy gap}
 \label{fig:dspe}   
 }
\end{figure}

Next, we consider a symmetric double-well potential, 
\begin{eqnarray}
V_0(x) =~~\lambda_0 x^4-\frac{1}{2} x^2 .
\end{eqnarray}
In this case, there is no well-defined perturbation theory. A standard
technique to get the energy gap $\Delta E$ is the dilute gas instanton
calculation which gives a result with the essential singularity, 
\begin{eqnarray}
\Delta E=2\sqrt{\frac{2\sqrt{2}}{\pi \lambda_0}}e^{-\frac{1}{3\sqrt{2}\lambda_0}}.
\end{eqnarray}
In NPRG evolution of the effective potential, the initial double-well
potential finally becomes a single well and the energy gap (mass)
arises (Fig.\ref{fig:dpote}).
This evolution is readily understandable considering that in one
space-time dimension  
the $Z_2$ symmetry does not break down due to the barrier
penetration, i.e. the quantum tunnelling. 
\par
\hspace{-6mm}The NPRG results are very good in the strong coupling region, where we
have no other method to compete (Fig.\ref{fig:dspe}). The perturbation can not be applied
in this double-well system and the dilute gas instanton does
not work at all, which is valid only in the very 
weak coupling region. Therefore NPRG method can be a
powerful tool for analysis of tunnelling at least in such
region.  
However, our NPRG results deviate from exact values as $\lambda _0 \to
0$, which corresponds to a very deep well. Because the $\beta$-function
becomes singular in this region, NPRG results become unreliable. We
consider that the cause of difficulty comes from the
approximation scheme that we now adopt. After all, the coupling regions 
where LPA W-H eqn. and the dilute gas instanton are valid respectively are separated
completely. There is only a cross
over region. In this sense, these two methods are complementary to
each other.  \par
\begin{figure}[htb]       
\hspace{8mm}
 \parbox{70mm}{
 \epsfxsize=70mm      
 \epsfysize=70mm
  \leavevmode
\epsfbox{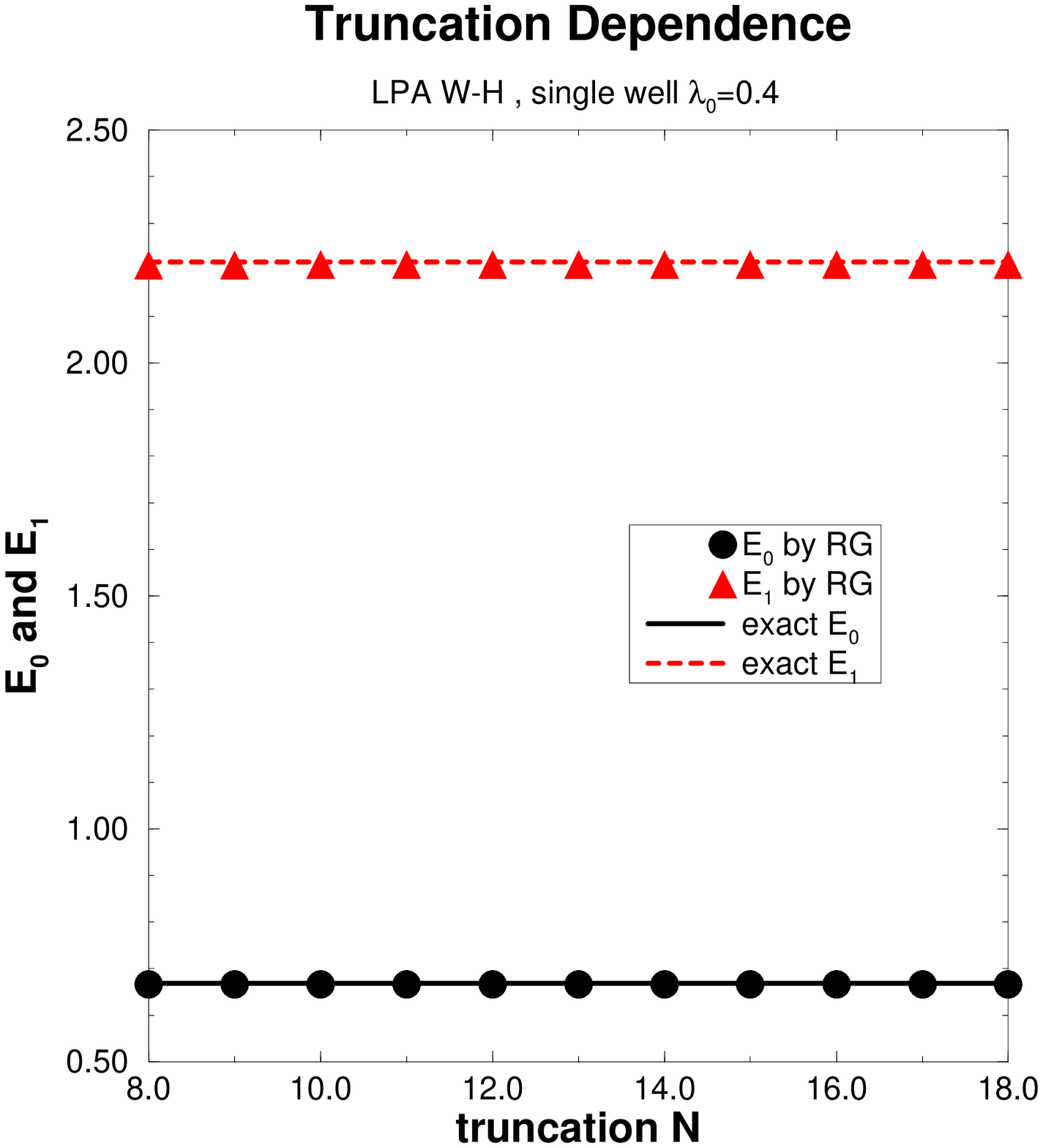}
\vspace{-10mm}
 \caption{Single well}
 \label{fig:strun}    
 }
\hspace{8mm} 
\parbox{70mm}{
 \epsfxsize=70mm      
 \epsfysize=70mm
 \leavevmode
\epsfbox{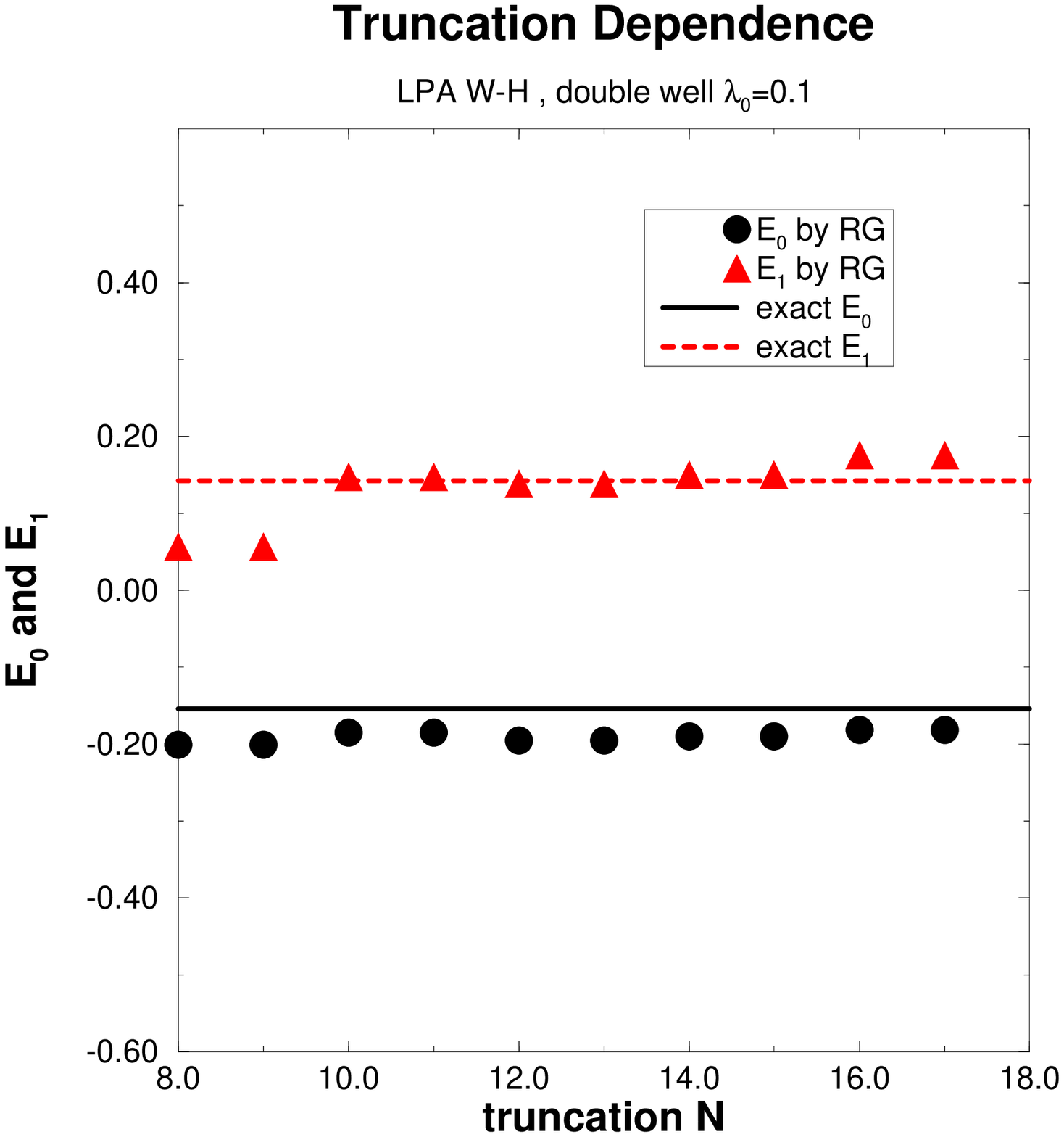}
\vspace{-10mm}
 \caption{Double well} 
 \label{fig:dtrun}     
 }
\end{figure}

Concerning the reliability of the results, we have to check the truncation dependence of physical quantities. In
a single-well case (Fig.\ref{fig:strun}), the results converge extremely well,
but in a double-well case, as $\lambda _0$ goes smaller (Fig.\ref{fig:dtrun}), the convergence becomes unclear. So in the small
$\lambda _0$ region, the effective couplings
at even $N=16$ are not suitable for physical quantities. We should always pay attention to 
the convergence with respect to $N$.\cite{ka} 
\par
\section{Supersymmetric quantum mechanics}
Finally we analyze the supersymmetric theory, where we can see the non-perturbative dynamics of the system
more clearly. We consider the Witten's toy model for dynamical SUSY
breaking\cite{wi}, whose Hamiltonian is represented as follows,
\begin{eqnarray}
\hat{H}=\frac12\left[\hat{P}^2+\hat{W}^2(\Phi)
				+\sigma_3\frac{d\hat{W}(\Phi)}{d\Phi}\right]=\left(\matrix{\!\!\!\frac{1}{2}\hat{P}^2+\hat{V}_{+}(\Phi)~~~~~~~0\cr ~~0~~~~~~~~~~\frac{1}{2}\hat{P}^2+\hat{V}_{-}(\Phi)}\right) ,
\end{eqnarray}
where $\hat{V}_{\pm}(\Phi)=\frac{1}{2}\hat{W}^2(\Phi)\pm
\frac{1}{2}\frac{d\hat{W}(\Phi)}{d\Phi}$ and $\hat{W}(\Phi)$ is called 
SUSY potential. We define super charges 
$\hat{Q}_1=\frac12(\sigma_1\hat{P}+\sigma_2\hat{W}(\Phi))$, 
$\hat{Q}_2=\frac12(\sigma_2\hat{P}-\sigma_1\hat{W}(\Phi))$ and the
Hamiltonian is written as
$\hat{H}=\{\hat{Q}_1,\hat{Q}_1\}=\{\hat{Q}_2,\hat{Q}_2\}$. This
assures that the vacuum energy is always non-negative, 
$E_0=\langle\Omega|\hat{H}|\Omega\rangle
     =2\left\Vert\hat{Q}_1|\Omega\rangle\right\Vert ^{2}
     =2\left\Vert\hat{Q}_2|\Omega\rangle\right\Vert ^{2}\geq0$, and we have 
     the criterion of SUSY `breaking',  
\begin{eqnarray}
E_0=0 \quad \Rightarrow
	&&\hat{Q}_1|\Omega\rangle=0 , \hat{Q}_2|\Omega\rangle=0 \qquad
	{\rm SUSY\quad unbroken} ,\\
E_0>0 \quad \Rightarrow
	&&\hat{Q}_1|\Omega\rangle\ne0 , \hat{Q}_2|\Omega\rangle\ne0
	\qquad {\rm SUSY\quad breaking} .
\end{eqnarray}
That is, the vacuum energy $E_0$ is the order parameter of SUSY
`breaking'. Furthermore, the perturbative corrections to $E_0$ are vanishing in any order 
of perturbation, which is known as the non-renormalization
theorem. Actually under the SUSY potential $W(\Phi)=g\Phi^2-\Phi$,
$V_{+}(\Phi)$ becomes,
\begin{eqnarray}
V_+(\Phi) =~~\frac{1}{2}g^2\Phi^4-g\Phi^3+\frac{1}{2}\Phi^2+g\Phi-\frac{1}{2},
\end{eqnarray}
and the perturbative corrections to energy spectrum are calculated as follows,  
\begin{eqnarray}
E_n=n+\frac38g^2[2n^2+2n+1]-\frac38g^2[10n^2+2n+1]
%-\frac{1}{32}g^4[34n^3+51n^2+59n+21]
+\cdots .
\end{eqnarray}
These corrections to $E_0$ are cancelled out in each order
of $g$, thus there is no perturbative corrections. Namely, non-vanishing $E_0$ is realized only by
non-perturbative effects caused by the essential singularity. \par 
\begin{figure}[htb] 
\hspace{8mm}
 \parbox{70mm}{
 \epsfxsize=70mm     
 \epsfysize=70mm
  \leavevmode
\epsfbox{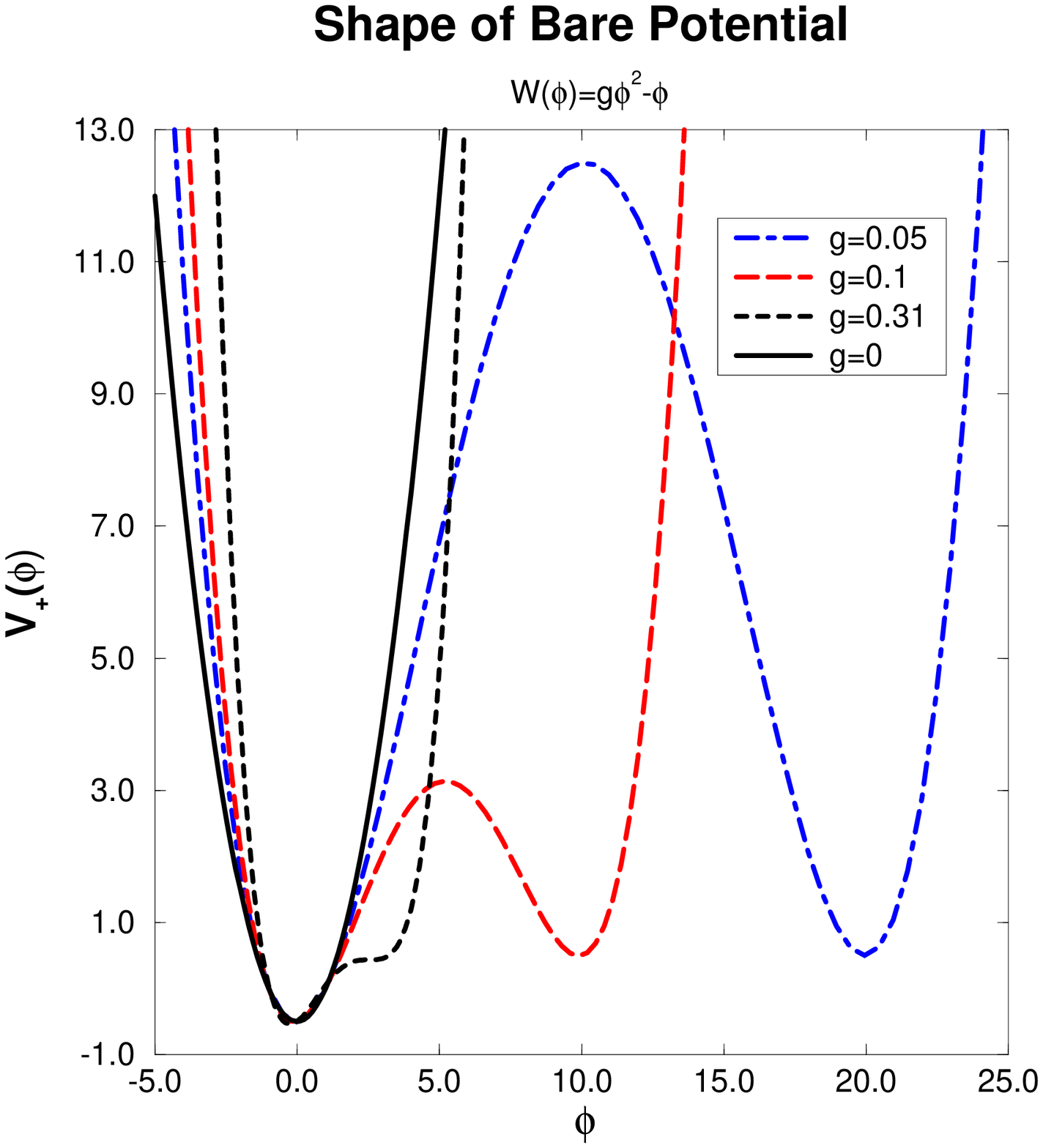}
\vspace{-10mm}
 \caption{Bare potentials}  
 \label{fig:bpote}        
 }
\hspace{8mm} 
\parbox{70mm}{
 \epsfxsize=70mm     
 \epsfysize=70mm
 \leavevmode
\epsfbox{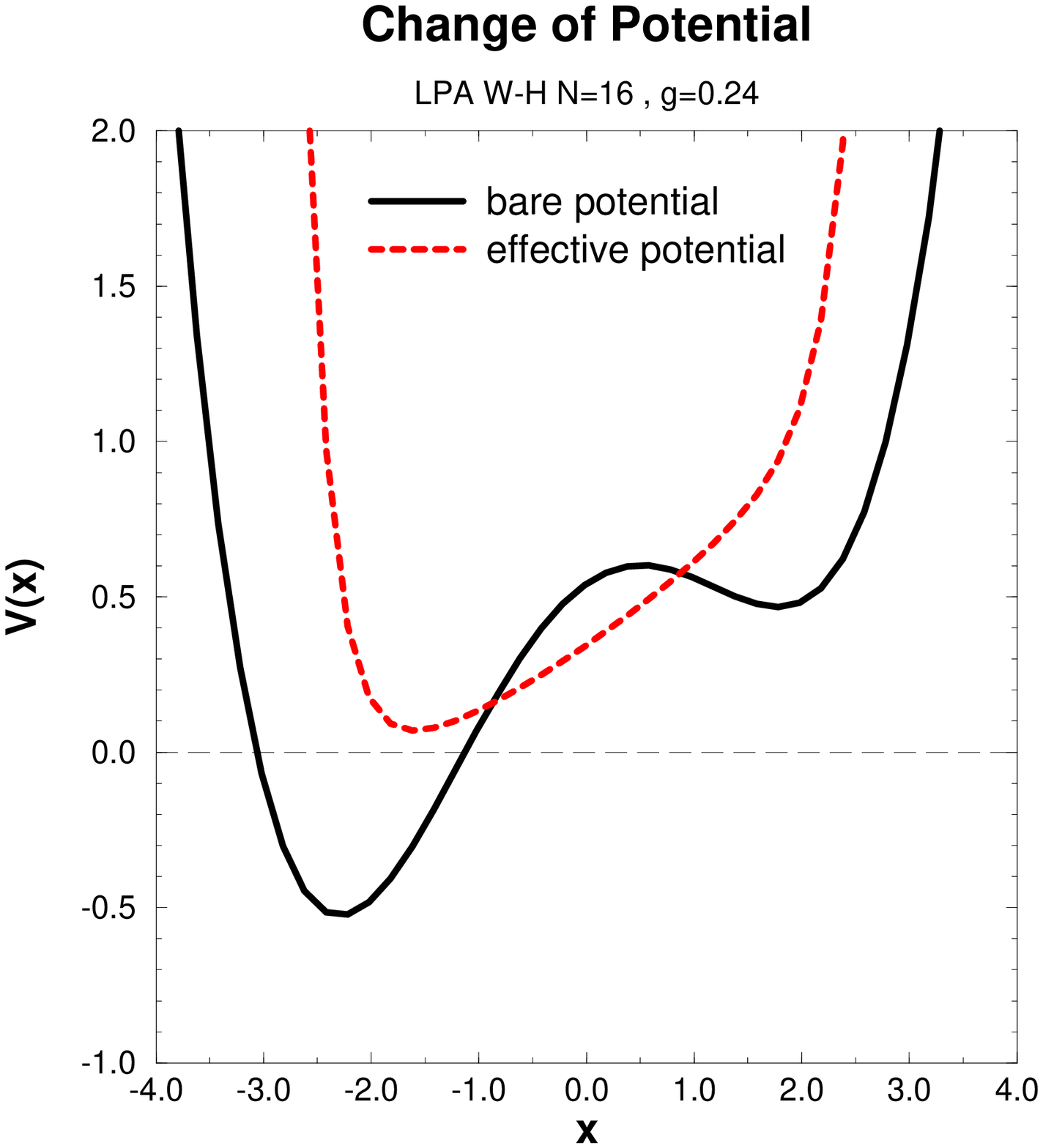}
\vspace{-10mm}
 \caption{Change of potential}
 \label{fig:epote}       
 }
\end{figure}
Now we analyze this system by our NPRG method. We calculate the effective potentials for a wide range of parameter
$g$. The case of vanishing $g$ corresponds to the harmonic oscillator and SUSY
does not break there. On the other hand, SUSY is dynamically broken at any
non-vanishing $g$. Note that at small $g$ the bare potential is an asymmetric
double-well, while at $g > \sqrt[4]{\frac{1}{108}}\simeq 0.31$ it is a single-well and quantum
tunnelling is irrelevant there (Fig.\ref{fig:bpote}). Figure
\ref{fig:epote} shows the result for
$g$=0.24, where the effective potential evolves into a convex one and
its minimum turns out to be positive. That is, our NPRG method
gives positive $E_0$ correctly and realizes the dynamical SUSY breaking.           
\begin{figure}[htb]
\hspace{8mm}
 \parbox{70mm}{
 \epsfxsize=70mm     
 \epsfysize=70mm
  \leavevmode
\epsfbox{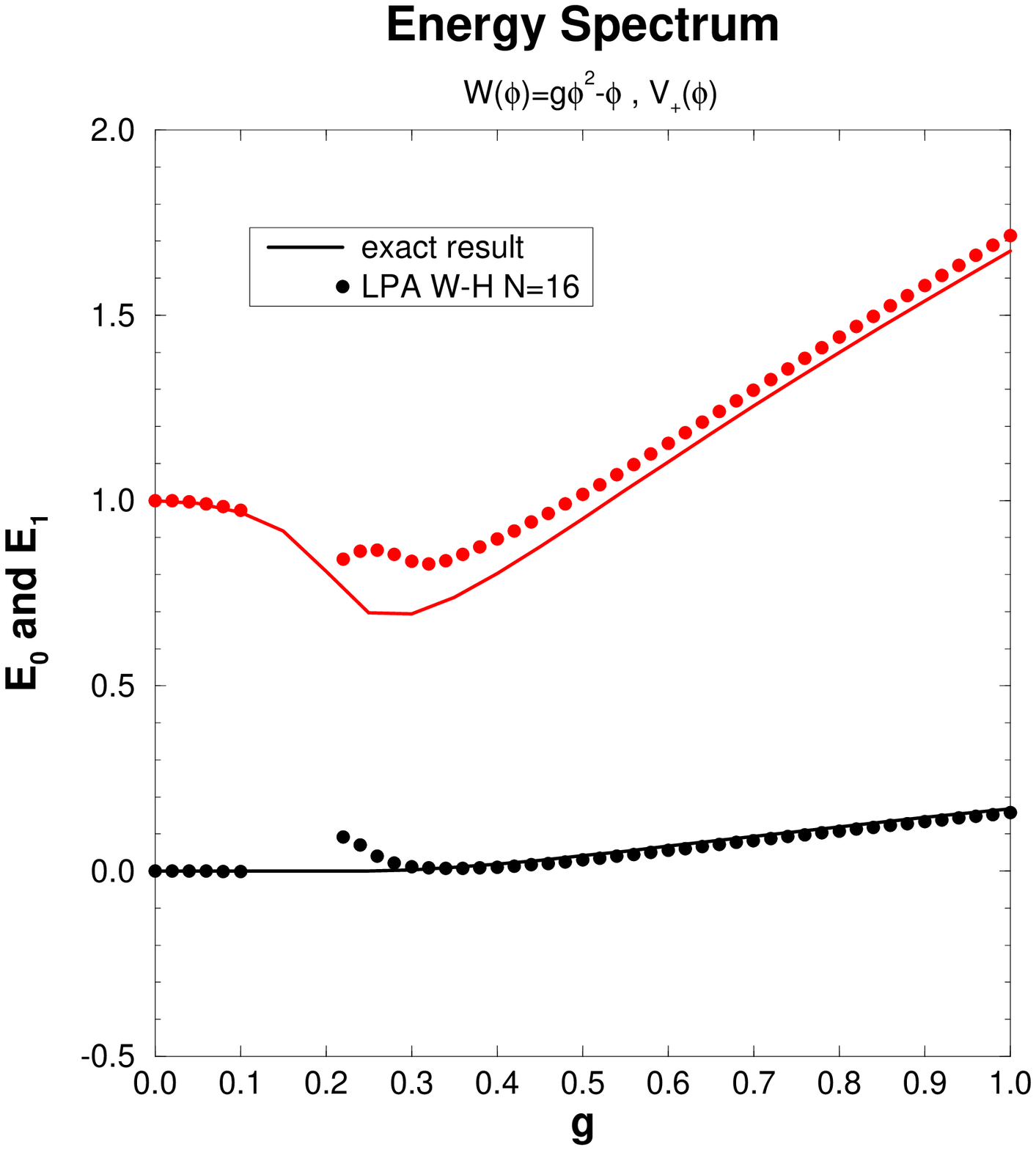}
\vspace{-10mm}
 \caption{Energy spectrum} 
 \label{fig:rspe}       
 }
\hspace{8mm} 
\parbox{70mm}{
 \epsfxsize=70mm      
 \epsfysize=70mm
 \leavevmode
\epsfbox{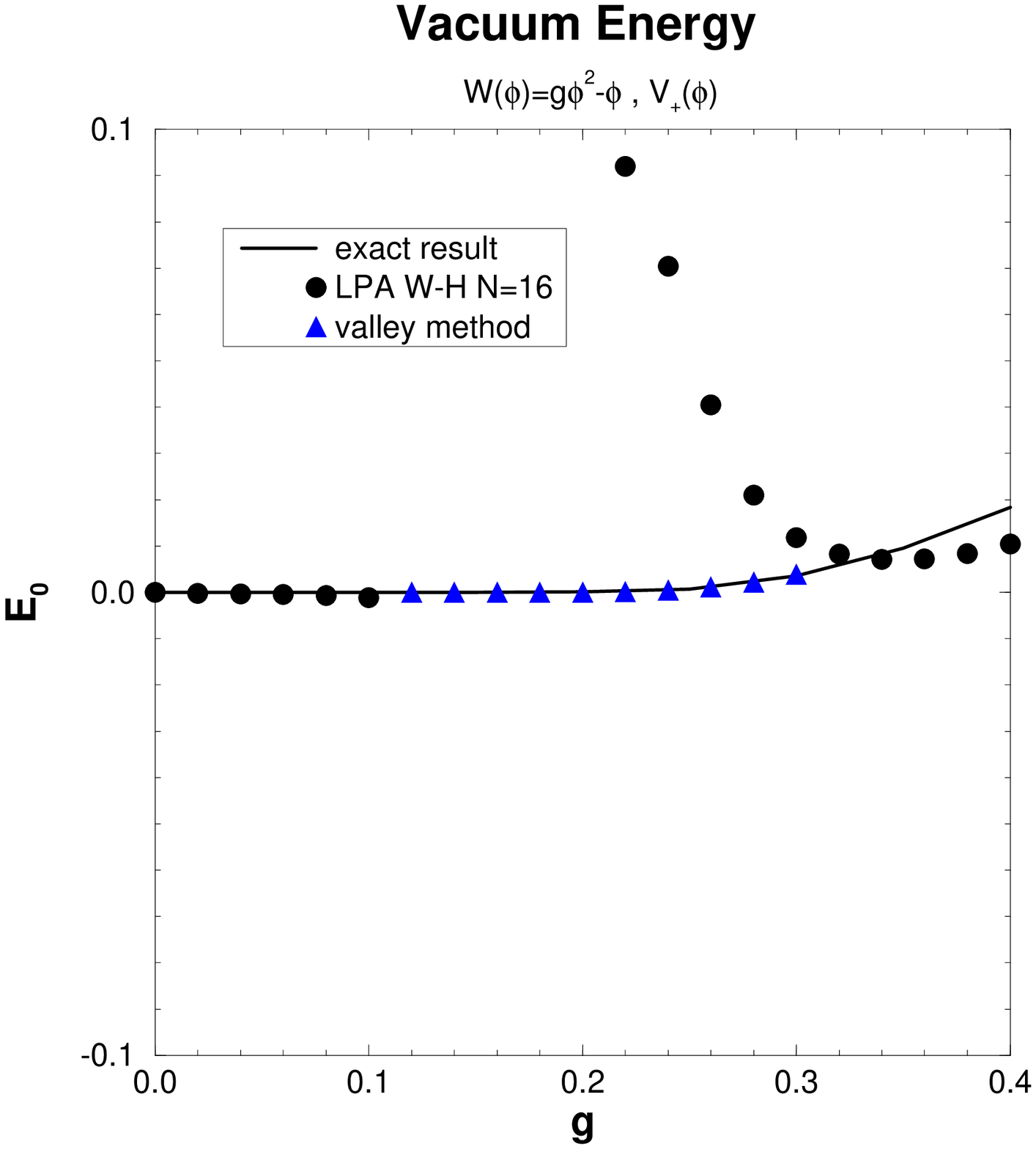}
\vspace{-10mm}
 \caption{RG and valley instanton }
 \label{fig:vspe}      
 }
\end{figure}
As is shown in Fig.\ref{fig:rspe}, NPRG results
 are excellent in the strong coupling region, but not in the region where the
 bare double-well potential becomes deep. In this region ($ 0.1<g<0.2
 $), we can not show any result because of large numerical errors,
 while the valley instanton method works very well as shown in
 Fig.\ref{fig:vspe}. The valley instanton is generalization of the instanton method based on
 the valley structure of the configuration space.\cite{ao1,ao2} Again, two methods are
 somehow complementary to each other.  
\section{Discussions}
The NPRG method, even in LPA, can evaluate the non-perturbative
quantities of the summation of all orders of the perturbative series
in a quantitatively good manner. As for the
non-perturbative quantities characterized by the essential
singularity, LPA W-H eqn. also works very well in the region where
the instanton and the perturbation break down, i.e. the strong
coupling region. However NPRG is not so effective in the weak coupling
region because of large numerical errors. To summarize, LPA W-H
eqn. and the (valley) instanton
play complementary roles to each other. We don't know the clear origin of difficulty 
which we encounter in our NPRG analysis. We suspect that the
derivative expansion does not fit in such a parameter
region. Anyway, we have to search for `better' approximations. \par
On the other hand, from a practical point of view, NPRG method can
be a good new tool for analysis of quantum tunnelling
at least in some parameter region. For this purpose, we also need to study in detail 
how to extract tunnelling physics from the effective potential and
the effective action.   
Those techniques may be applied to models in quantum field
theories\cite{st} and in more complex systems. Especially quantum tunnelling
with multi-degrees of freedom represented by dissipation\cite{cj} is a very interesting subject to be attacked
by NPRG method. 
%%%%%%%%%%%%%%%%%%%%%%%%
\section*{Acknowledgments}
K.-I.Aoki and H.Terao are partially supported by the Grant-in Aid for
Scientific Research \\ (\#09874061, \#09226212, \#09246212, \#08640361) from the
Ministry of Education, Science \\and Culture.
%%%%%%%%%%%%%%%%%%%%%%%%

\end{document}